\documentclass[preprint,review,12pt]{elsarticle}

 \usepackage{graphics}

\usepackage{amssymb}

\begin{document}

\begin{frontmatter}



\title{A Parametric Study Examining the Effects of Reshock in RMI}

\author{Victoria Suponitsky}
\ead{victoria.suponitsky@generalfusion.com}s
\author{Aaron Froese}
\author{Sandra Barsky}
\address{General Fusion Inc., 108-3680 Bonneville Place,
Burnaby, BC V3N 4T5,Canada}

\begin{abstract}

%
%

The compression of a cylindrical gas bubble by an imploding molten
lead (\emph{Pb}) shell may be accompanied by the development of
the Richtmyer-Meshkov (RM) instability at the liquid-gas interface
due to the initial imperfection of the interface.
A converging pressure wave impinging upon the interface causes a
shell of liquid to detach and continue to travel inwards,
compressing the gas bubble. The efficiency of compression and
collapse evolution can be affected by development of the RM
instability. Investigations have been performed in the regime of
extreme Atwood number $A=-1$ with the additional complexity of
modeling liquid cavitation in the working fluid. Simulations have
been performed using the open source CFD software OpenFOAM on a
set of parameters relevant to the prototype compression system
under development at General Fusion Inc. for use as a Magnetized
Target Fusion (MTF) driver.

After validating the numerical setup in planar geometry,
simulations have been carried out in 2D cylindrical geometry for
both initially smooth and perturbed interfaces. Where possible,
results have been validated against existing theoretical models
and good agreement has been found. While our main focus is on the
effects of initial perturbation amplitude and azimuthal mode
number,  we also address differences between this problem and
those usually considered, such as RM instability at an interface
between two gases with a moderate density ratio. One important
difference is the formation of narrow molten lead jets rapidly
propagating inwards during the final stages of the collapse. Jet
behaviour has been observed for a range of azimuthal mode numbers
and perturbation amplitudes.

\end{abstract}


\end{frontmatter}


\section{Introduction}
\label{sec:intro}

When the interface between two fluids of different densities is
subjected to rapid acceleration, e.g. by a shock passing through
the interface, perturbations present at the interface prior to the
passage of the wave grow with time. This phenomenon is known as
the Richtmyer-Meshkov instability (RM)~\cite{richtmyer, meshkov},
and has been extensively studied in the last couple of decades,
mainly in the field of astrophysics.
Recently, a renewed interest in RM was triggered by innovations in
magnetized target fusion (MTF).

In this paper we study the cylindrical collapse of a gas cavity by
an imploding liquid shell,
where the development of interface instabilities is known to
affect the compression efficiency. In this study we focus on the
RM instability~\cite{richtmyer, meshkov}, which is the first
instability to develop on the liquid-gas interface during
collapse. The MTF driver in development at General Fusion Inc.
will compress a plasma-filled cylindrical cavity by using
pneumatic pistons to initiate a converging pressure wave in molten
lead (\emph{Pb}). As the pressure wave reaches the liquid-plasma
interface, the interface undergoes rapid acceleration and travels
toward the center, compressing the plasma cavity.  In our
prototype device, the cavity is instead filled with argon gas.
Perturbations existing at the liquid-gas interface prior to the
passage of the pressure wave may seed the development of
hydrodynamic instabilities. In this study, we concentrate on the
parameter space relevant to our prototype device so as to define
requirements for the smoothness of the initial liquid-gas
interface for efficient compression.

The growth characteristics of initially small-amplitude sinusoidal
perturbations can be divided into two regimes: (i) the linear
regime, in which the contribution of nonlinear effects is
negligible and evolution of the disturbance can be adequately
described by the linearized equations, and (ii) the nonlinear
regime, in which the perturbation growth decreases and finally
saturates due to nonlinear effects. In the linear regime, initial
perturbation growth can be reasonably predicted by the simple
impulse models~\cite{richtmyer, meshkov} as $\dot {h}=h_{0}A
\Delta U$, where $h_{0}$ is the initial perturbation amplitude,
$\Delta U$ is the difference in the velocity of the interface
before and after the passage of the shock wave, and $A$ is the
Atwood number defined as
$A=(\rho_{2}-\rho_{1})/(\rho_{2}+\rho_{1})$, where $\rho_{1}$ and
$\rho_{2}$ are the fluid densities.

In literature, a characteristic pattern of RM instability is
usually described in terms of fingers of one fluid penetrating
into another. A finger of light fluid poking into heavy fluid is
usually called a `bubble', and that of heavy into light is called
a `spike'. Bubbles and spikes grow at the same rate during the
linear stage. However during the nonlinear stage, spikes undergo
acceleration whereas bubbles tend to stagnate. The disparity in
growth rates becomes more prominent at high Atwood numbers ($A
\approx 1$), see e.g. Dimonte and Ramaprabhu~\cite{Dimonte}, and
in this regime  a vast majority of the existing models also
perform poorly.

Once the passing shock wave and the interface begin to interact,
the evolution of the initially perturbed interface can be
explained in terms of vorticity deposition. If the interface is
perturbed, the pressure gradient of the shock is misaligned with
the density gradient across the interface. This results in
generation of the baroclinic vorticity through the term $\nabla
\rho \times \nabla p$ in the vorticity equation. The sign of the
generated vorticity (clockwise or counter clockwise) depends on
the sign of the Atwood number, i.e. whether the shock travels from
light fluid to heavy or vice versa~\cite{brouillette,Zabusky}. As
such, the initial perturbation may grow monotonically or first
decrease and then grow in the opposite direction, a phenomenon
known as phase inversion~\cite{graham,Zabusky}.

Most of the work to date on the RM instability has been carried
out in rectangular geometry, with fluids modeled as ideal gases,
and at moderate Atwood numbers ($|A| \approx 0.5 - 0.8$). A lot of
effort was put into understanding the underlying physics,
developing models describing nonlinear stages of disturbance
evolution, and investigating the effects of compressibility,
sensitivity to the initial conditions, and turbulent mixing,
e.g.~\cite{yang,holmes,brouillette,latini,Dimonte,Nishihara,Thornber,Cohen}.
Recently, there has been an increase in the number of works
describing the RM instability in converging
geometries~\cite{graham,fincke,tian,krechetnikov,Mikaelian,lombardini},
which are more relevant to fusion. The situation in converging
geometries is more complex than the planar case, because the
trajectories of bubbles and spikes are no longer parallel as the
interface moves in the radial direction. The evolution of small
amplitude perturbations in cylindrical geometry was investigated
by Mikaelian~\cite{Mikaelian} for the case of pure azimuthal
perturbations.  Lombardini~\cite{lombardini} extended the
analysis~\cite{Mikaelian} to also account for axial perturbations.

In converging geometries a fair bit of attention is devoted to the
secondary effects, such as so-called `reshock'
~\cite{graham,tian}; in planar geometry `reshock' has been studied
by~\cite{latini}. When a shock wave strikes the interface between
two fluids, it is partially transmitted into the second fluid and
partially reflected. The transmission ratio depends on the
acoustic impedance of each fluid, defined as $z=\rho c$, where
$\rho$ and $c$ are the density and sound speed of the fluid,
respectively. In a converging geometry the transmitted part of the
shock travels to the origin, which acts as a singular point, and
then bounces back to hit the interface again affecting the
perturbation growth; this phenomenon is called `reshock'.

One of the aspects of the RM instability that has been given
little attention until recently is the regime of high Atwood
number ($A \approx \pm 1$). This situation occurs when a shock
wave passes, for example, between a liquid and a gas. In this
case, at least one of the fluids cannot be described as an ideal
gas and other equations of state must be considered. If one of the
fluids is a liquid and the shock reflects off the interface,
cavitation can occur when the pressure falls below the tensile
strength of the liquid, further increasing the complexity of the
problem. A recent numerical study of Ward and Pullin~\cite{Ward}
looks into the role that the equation of state has on RM
instability growth in a planar geometry. An experimental study by
Buttler \emph{et al.}~\cite{Buttler} investigates the RM
instability at metal-vacuum interfaces in planar geometry. Their
focus was on developing an ejecta source term model that links to
the surface perturbations of shocked materials. The main
assumption of their model is that ejecta formation at a
metal-vacuum interface can be viewed as a special limiting case of
the RM instability.

This numerical work focuses on the RM instability at a liquid-gas
interface during a heavy-to-light implosion in cylindrical
geometry. In this case, we have liquid $Pb$ surrounding a
cylindrical cavity of air. The pressure pulse originates in the
liquid and converges toward the liquid-gas interface. When the
pressure pulse reaches the interface, it is almost entirely
reflected because of the severe mismatch between the acoustic
impedance of liquid $Pb$ and that of air. In this configuration,
the reflected wave is a rarefaction wave that subjects the liquid
to tension, which may cause cavitation.

The rest of this article is organized as follows. The problem
statement and numerical method are described in $\S
\ref{sec:method}$. The results of pulse propagation in liquid $Pb$
and collapse of the unperturbed cylindrical gas cavity are given
in $\S \ref{sec:noPert}$.  The effects of perturbation amplitude
and azimuthal mode number on the RM instability growth rate are
studied in $\S \ref{sec:RM}$. Finally, all results are summarized
in $\S \ref{sec:Sum}$.


\section{Problem Statement and Validation}
\label{sec:method}

\subsection{Numerical Setup}

Simulations are performed using the open source CFD code
OpenFOAM~\cite{openfoam}. A compressible multiphase solver
\emph{`compressibleInterFoam'} is used for these simulations. This
solver implements the Volume of Fluid (VoF) method for interface
tracking and is suitable for the simulation of two compressible
immiscible isothermal fluids. A barotropic equation of state is
used to relate pressure and density for each phase:
\begin{equation}
\rho_i=\rho_{0_i}+\psi_i P, \label{eq:state}
\end{equation}
where $\psi_i=1/c_i^2$ is the compressibility and $c_i$ is the
speed of sound for phase $i$. For a gas (compressible) phase  the
nominal density $\rho_0$ in Eq.~\ref{eq:state} is set to zero.
This results in an ideal gas equation of state for an isothermal
fluid. For a liquid (low compressibility phase) $\rho_0$ is set to
the nominal density of the liquid under normal conditions. As such
the fluid density remains essentially constant unless the liquid
is subjected to very high pressures. Similar results can also be
obtained using the Tait~\cite{Tait} equation of state for a liquid
phase.

\begin{figure}[tbp]
\begin{center}
\includegraphics[width=0.8\linewidth]{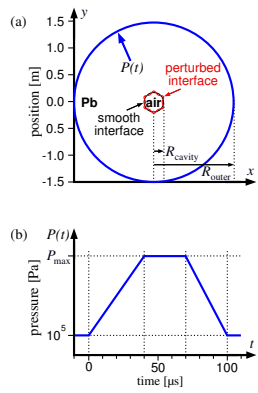}
\end{center}
\caption{(a) Numerical setup for 2D simulations in a cylindrical
geometry; (b) Typical shape of the pressure pulse imposed on the
outer boundary.} \label{InitialSetup}
\end{figure}

Simulations are carried out in 2D cylindrical geometry. A
schematic of the numerical setup is shown in
Fig.~\ref{InitialSetup}(a). The initial radius of the gas cavity
and initial position of the interface is $R_{0}=0.2$ m. The outer
boundary is at $R_\mathrm{outer}=1.5$ m, where a pressure pulse is
imposed as a time-dependent pressure boundary condition $P(t)$. A
typical pressure pulse is shown in Fig.~\ref{InitialSetup}(b). The
maximum pressure and duration of the pulse are chosen to reflect
the parameters of the prototype device, $P_\mathrm{max}=1.5$ GPa
and $T_\mathrm{pulse} \approx 100$ $\mu$s in most simulations. A
zero-gradient boundary condition is set for the velocity at the
outer boundary allowing some mass influx into the domain. In most
simulations a small central portion of the computational domain
with radius $r < 1$ cm has been excluded from the calculations to
speed up the computations. The effect of excluding this central
part was found to be negligible for the purposes of this work. The
inner boundary uses an outflow boundary condition so that gas can
`escape' from the domain during the collapse. Both fluids are
initially at rest at atmospheric pressure.

Simulations are carried out for both initially smooth and
sinusoidally perturbed interfaces. The results for the unperturbed
case are first validated against existing models and then used as
a baseline of comparison for the growth of bubbles and spikes in
perturbed interface runs.

The initial sinusoidal perturbation is defined by its initial
amplitude $h_{0}$ and azimuthal mode wavenumber $n$, with a
corresponding perturbation wavelength of $\lambda_{0}=2\pi
R_{0}/n$.

The perturbation amplitude can be normalized relative to
its wavelength, $h_{0}/\lambda_{0}$, as is done in planar
geometry, or relative to the initial radius of the interface
$h_{0}/R_{0}$.

Simulations with a coarse grid resolution in the radial direction
($\hat{r}$) are performed over the full azimuthal domain
(Fig.~\ref{InitialSetup}a), while finer grid resolution runs are
carried out over a restricted azimuthal angle
$\theta_\mathrm{segment}$ (with periodic boundary conditions in
the azimuthal direction) to speed up the calculations. The
specific choice of $\theta_\mathrm{segment}$ depends on the
azimuthal mode number $n$ of the perturbation used. The number of
grid points in the radial direction is $N_r=2800$ and $N_r=11200$
for the coarse and fine resolution runs, respectively. The grid
spacing is uniform for $r<R_{0}$ with $dr=2.375 \times 10^{-4}$ m
and $dr=5.9375 \times 10^{-5}$ m for the coarse and fine grids,
respectively. The smallest perturbation amplitude used in these
simulations is $h_{0}=0.001$ m. This results in 17 fine grid
points across the initial perturbation in the radial direction.
The number of grid points per perturbation wavelength is set to
$N_{\theta}=135$ in most simulations, although this is reduced to
$N_{\theta}=55$ for high azimuthal mode perturbations.

 Simulations are performed for an implosion of
 molten lead $Pb$ into air with the fluid properties
$\rho_{Pb}=10000$ kg/m$^3$, $c_{Pb}=2000$ m/s; $\rho_\mathrm{air}=1$
kg/m$^3$, $c_\mathrm{air}=316$ m/s. The corresponding acoustic impedances
are $z_{Pb}=2\times10^7$ Rayl and $z_\mathrm{air}=316$ Rayl
and the Atwood number is
$A=(\rho_\mathrm{air}-\rho_{Pb})/(\rho_\mathrm{air}+\rho_{Pb})=-0.9998 \approx -1$.

\subsection{Validation Test}

The numerical method and grid convergence are first tested in
planar geometry on a set of parameters similar to those used in
cylindrical geometry. The length of the computational domain in
the streamwise direction ($\hat{x}$, normal to the interface) is
$L_{x}=1.5$ m with the interface located at
$X_\mathrm{interface}=1.3$ m. A pressure pulse $P(t)$ is
prescribed at the inflow boundary at $X=0$ and an outflow
(zero-gradient) boundary condition is used at $X=1.5$ m. The
initial pressure pulse amplitude is $P_{max}=3.6$ GPa which
roughly corresponds to the expected pressure at the interface in
cylindrical geometry when the initial pulse amplitude is
$P_{max}=1.5$ GPa (pressure is amplified due to the convergence).
In the normal direction ($\hat{y}$, parallel to the surface of the
interface) the length of the computational domain is one
wavelength of a mode $n=6$ perturbation in the case of cylindrical
geometry, $L_{y}=2\pi R_{0}/n=0.2094395102$ m. Periodic boundary
conditions are imposed in the normal direction and the number of
grid points set to be $N_{y}=135$, corresponding to the $n=6$ case
in cylindrical geometry. Simulations are performed for three
different grid resolutions in the streamwise direction. The grid
spacing is uniform for $1.3$ m $\le X \le 1.5$ m and equal to
$\Delta x =2.375 \times 10^{-4}$, $1.1875 \times 10^{-4}$ and
$5.9375 \times 10^{-5}$ m
for grids with increasing resolution. The total number of grid
points in the streamwise direction is correspondingly
$N_{x}=2800$, $5600$ and $11200$.

\begin{figure}[tbp]
\begin{center}
\includegraphics[width=0.5\linewidth, angle=0]{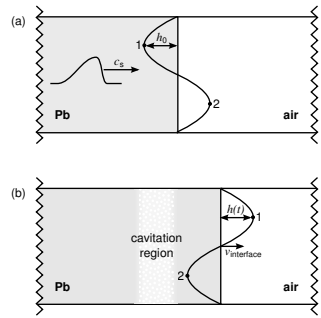}
\end{center}
\caption{Schematic of a test case in planar geometry. (a) Pressure
pulse propages through $Pb$ prior to hitting the $Pb$-air
interface. (b) Some time after the pressure pulse hits the
$Pb$-air interface.} \label{RectangularSetup}
\end{figure}

A schematic of the flow pattern in the planar case is shown in
Fig.~\ref{RectangularSetup}. Part (a) of the figure shows
propagation of the pressure pulse through $Pb$ prior to hitting
the interface and part (b) illustrates the flow pattern some time
after the pressure pulse hit the interface. The pressure pulse
gets reflected from the interface as a rarefaction wave. This puts
the liquid $Pb$ into tension and initiates cavitation behind the
interface (Fig.~\ref{RectangularSetup}b).

In planar geometry, an initially smooth interface is expected to
move with constant velocity after interacting with the pressure
pulse. In our case, the interface velocity $V_\mathrm{interface}$
can  be  approximated by assuming two fluids with a very large
impedance ratio (see $\S 4.8$ and $\S 4.9$ in ~\cite{Thompson}),
such that,
\begin{equation}
V_\mathrm{interface} \approx 2  V_p \,, \label{eq:VelInt}
\end{equation}
where $V_p$ is a particle velocity given by,
\begin{equation}
V_p=\frac{P_\mathrm{max}-P_{0}}{\rho_{Pb}c_{Pb}} \approx
\frac{P_\mathrm{max}}{\rho_{Pb}c_{Pb}}. \label{eq:partVel}
\end{equation}
The maximum pressure $P_\mathrm{max}$ and density $\rho_{Pb}$ of
the pulse are taken just before it hits the interface.  The
ambient pressure is $P_{0}=1\times 10^{5}$ Pa.

\begin{figure}[tbp]
\begin{center}
\includegraphics[width=0.7\linewidth, angle=0]{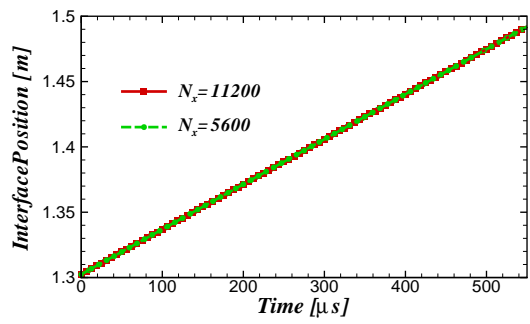}
\caption{\label{RecGridRef} Planar geometry: Evolution of the
initially smooth interface for two grid resolutions. Pressure
pulse is of an infinite length and amplitude $P_\mathrm{max}=3.6$
GPa.}
\end{center}
\end{figure}

The evolution of an initially smooth interface after it has been
accelerated by a linearly ramped pressure pulse of infinite length
is shown in Fig.~\ref{RecGridRef} for two different grid
resolutions. One can see that the grid resolution is sufficient to
obtain well converged results. For our parameters, $P_\mathrm{max}
\approx 3.6$ GPa, $\rho_{Pb} \approx 10500$ kg/m$^3$ (there is a
slight change in $Pb$ density as the pulse propagates through it
due to compressibility) and $c_{Pb} \approx 2000$ m/s, the
interface velocity predicted by Eq.~\ref{eq:VelInt} is
$V_\mathrm{interface}=343$ m/s.  The numerically calculated
velocity (slope of the curve) is $U_\mathrm{interface}=344.7$ m/s,
which deviates from the theoretical value by less than 0.5\%.

An initially perturbed interface is also tested in planar
geometry. The initial perturbation amplitude is set to $h_{0}=2$
mm (Fig.~\ref{RectangularSetup}) and the perturbation wavelength
$\lambda_{0}$ is equal to the length of the computational domain
in the normal direction $\hat{y}$ leading to
$h_{0}/\lambda_{0}=9.55 \times 10^{-3}$.

In our analysis we follow the extrema points of the perturbations,
marked by points $1$ and $2$ in Fig.~\ref{RectangularSetup}, which
we label, respectively, as bubbles and spikes throughout the
entire simulation. It is important to note that during the first
stage of the evolution phase inversion~\cite{graham} occurs. This
means that the perturbation which is initially a spike, i.e. heavy
fluid surrounded by light fluid reverses to become a bubble; and
vice-versa. As such our label `spike' corresponds to a finger of a
heavy fluid surrounded by light fluid once the phase inversion has
occurred, whereas at early stages it is a finger of light fluid
surrounded by heavy. The converse applies to bubbles.

\begin{figure}[tbp]
\begin{center}
\includegraphics[width=0.9\linewidth, angle=0]{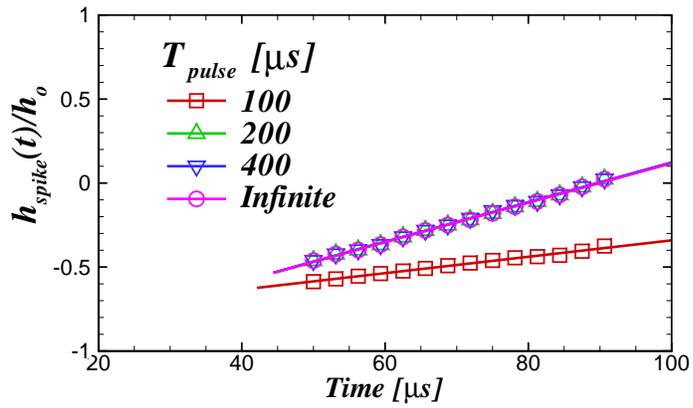}
\caption{\label{RecGrowthRates} Planar geometry: Early evolution
of the normalized spike amplitude for pressure pulses with various
lengths. Maximum pulse pressure is $P_\mathrm{max}=3.6$ GPa and
initial perturbation amplitude is $h_{0}/\lambda=9.55\times
10^{-3}$ .}
\end{center}
\end{figure}

Early evolution of the normalized spike amplitude is shown in
Fig.~\ref{RecGrowthRates} for pressure pulses of different
duration, each with a maximum pressure of $P_\mathrm{max}=3.6$ GPa
and modeled at the finest grid resolution. The spike amplitude has
been calculated as the difference between the interface position
at point $1$ for the case of a perturbed interface and the
coincidental position of the initially unperturbed interface. One
can see that for longer pulses ($T_\mathrm{pulse} \ge 200$ $\mu$s)
the pulse length has no effect on the spike amplitude. However, if
the pulse length falls below some threshold, the spikes' amplitude
growth slows down, clearly seen by  comparing results in
Fig.~\ref{RecGrowthRates} for the shortest pulse
$T_\mathrm{pulse}=100$ $\mu$s (red line) with those obtained for
longer pulses.

Initial disturbance growth rates (indicated by the slope of the
curves in Fig.~\ref{RecGrowthRates}) together with the growth rate
predicted by the Richtmyer impulsive model~\cite{richtmyer} (given
below by Eq.~\ref{eq:Richt}) for our set of parameters are listed
in Table~\ref{tab:RecRes}.
\begin{equation}
\dot{h}_\mathrm{planar}=h_{0}^+ k  A^+ \Delta U \label{eq:Richt},
\end{equation}
where $k=2\pi/\lambda$ is the wave number of the perturbation,
$h_{0}^+$ and $A^+$ are the initial post-shock amplitude
and Atwood number, and $\Delta U$ is the velocity jump at the
interface following passage of the shock. In our validation case
we use $k=30$, $h_{0}^+=2$ mm, $A^+=1$, and
$\Delta U =V_\mathrm{interface}=345$ m/s, resulting in
$\dot{h}\approx 20.7$ m/s (see Table~\ref{tab:RecRes}).

Our results for longer pulses agree well with the Richtmyer
impulsive model, while for a shorter pulse the growth rate is
lower.

\section{Results}
\label{sec:results}
\subsection{Pulse propagation and gas cavity collapse: unperturbed
interface} \label{sec:noPert}

In this section we study the collapse of an initially unperturbed
gas cavity in 2D cylindrical geometry. The numerical results are
validated against existing theoretical models and also used as a
baseline for calculations of the perturbation growth for the runs
with initially perturbed interfaces.

\begin{figure}[tbp]
\begin{center}
\includegraphics[width=1.0\linewidth]{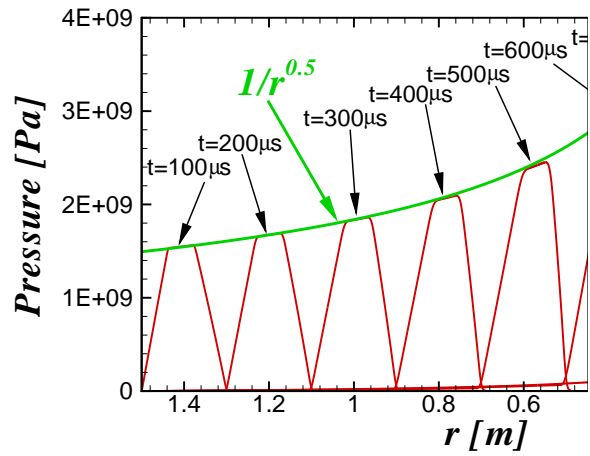}
\caption{\label{pulseProp} Pressure pulse propagation from the
outer boundary at  $r=1.5$ m  towards the interface at $r=0.2$ m
in cylindrical geometry.  Initial pulse amplitude is
$P_\mathrm{max}=1.5$ GPa and pulse duration is
$T_\mathrm{pulse}=100$ $\mu$s. The pulse propagates from left to
right.}
\end{center}
\end{figure}

Propagation of a pressure pulse through the liquid $Pb$ from the
outer boundary towards the interface in a cylindrical geometry is
shown in Figure~\ref{pulseProp}. The pressure pulse has a maximum
initial amplitude of $P_\mathrm{max}=3.6$ GPa and duration
$T_\mathrm{pulse}=100$ $\mu$s. It can be seen that the pressure
pulse is amplified as it cylindrically converges in the $Pb$. This
amplification is in excellent agreement with the theoretical
prediction for a small amplitude (linear) pulse that is $P \sim
1/\sqrt{r}$ in cylindrical geometry~\cite{Landau}. (The pulse is
expected to exhibit linear behaviour when the particle velocity is
much less than the sound speed.) As the pulse approaches the
interface and pressure becomes higher, nonlinearity starts to
manifest itself by a steepening of the pulse front and a slight
deviation of the amplitude from the theoretical curve. For the
current set of parameters (near the interface: $P_\mathrm{max}
\approx 3.6$ GPa, $\rho_{Pb} \approx 10500$ kg/m$^3$ and $c_{Pb}
\approx 2000$ m/s) the particle velocity of the pulse near the
interface can be roughly estimated as $V_{p}\approx 171$ m/s by
Eq.~\ref{eq:partVel}. This velocity is still relatively small (but
not negligible) compared to the speed of sound ($171$ m/s compared
to $2000$ m/s). Thus the pulse exhibits predominantly linear
behaviour as it propagates through the $Pb$, although small
nonlinear effects become noticeable near the interface. The time
taken for the pulse to reach the interface $1.3$ m away is
$t_\mathrm{propagation}=650$ $\mu$s, which agrees with the
prescribed speed of sound $c_{Pb}=2000$ m/s. In subsequent
results, time is defined relative to the moment the pressure pulse
reaches the interface, such that $t=t_\mathrm{sim} -
t_\mathrm{propagation}$.

\begin{figure}[tbp]
\begin{center}
\includegraphics[width=1.0\linewidth]{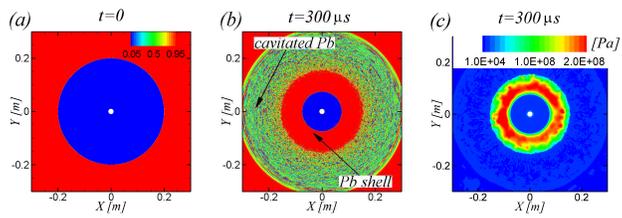}
\end{center}
\caption{\label{NDCollapseField} Typical structure of the flow
field during compression of initially unperturbed gas cavity by
pressure pulse with $T_\mathrm{pulse}=100$ $\mu$s and
$P_\mathrm{max}=1.5$ GPa. (a) Volume of fluid (VoF) contours when
the pressure pulse hits the interface; (b) Volume of fluid (VoF)
and (c) pressure contours when the cavity has partially collapsed
at $t=300$ $\mu$s.}
\end{figure}

A typical structure of the flow field during the collapse of an
initially unperturbed cylindrical cavity is shown in
Fig.~\ref{NDCollapseField} for a pressure pulse of duration
$T_\mathrm{pulse}=100$ $\mu$s and maximum pressure
$P_\mathrm{max}=1.5$ GPa. Parts (a) and (b) of the figure show
volume of fluid (VoF) contours when the pressure pulse strikes the
interface at $t=0$ and when the cavity has partially collapsed at
$t=300$ $\mu$s, respectively. Part (c) shows the corresponding
pressure contours at $t=300$ $\mu$s. It is worth reiterating that
the imploding material is liquid $Pb$ with an acoustic impedance
much larger than the air in the cavity. Therefore, the pressure
pulse is almost completely reflected back into the $Pb$ as a
rarefaction wave. The molten lead is then subjected to tension
which causes it to cavitate. It is apparent in
Fig.~\ref{NDCollapseField}b that a $Pb$ shell is formed as a
result of interaction between the pressure pulse and the
liquid-gas interface. As the shell moves inwards, a cavitation
region forms behind it, separating it from the rest of the molten
lead. The pressure contours in Fig.~\ref{NDCollapseField}c show
that the $Pb$ shell becomes pressurized as it converges, while the
pressure in the cavitation region falls to the minimum allowed by
the numerical setup.

\begin{figure}[tbp]
\begin{center}
\includegraphics[width=1.0\linewidth]{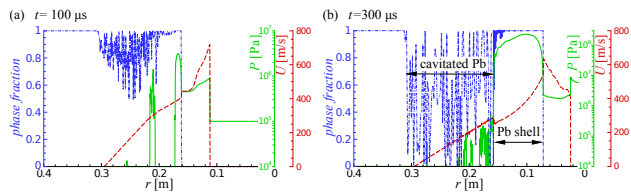}
\end{center}
\caption{\label{NDCollapseProfiles} Typical structure of the flow
field during compression of initially unperturbed gas cavity by
pressure pulse with $T_\mathrm{pulse}=100$ $\mu$s and $P_\mathrm{max}=1.5$
GPa. Radial profiles of VoF (blue dash-dot line),
pressure (green solid line), and velocity (red broken line) during
the collapse. (a) $t=100$ $\mu$s and (b) $t=300$ $\mu$s.}
\end{figure}

Radial profiles of the pressure, velocity and VoF at two different
instances during the collapse  ($t=100$ $\mu$s and $t=300$ $\mu$s)
are shown in Fig.~\ref{NDCollapseProfiles}. We mainly focus on the
behaviour of the molten lead as dynamics of the gas bubble has a
very little effect on the liquid $Pb$ until the very late stages
of the collapse. VoF profiles clearly show the location of the
liquid-gas interface and growth of the cavitation region as the
interface progresses inwards. Also evident is the increase in the
thickness of the $Pb$ shell as it converges during the collapse
process. From the pressure profiles we can see that the shell is
pressurized as it moves toward the center. The pressure in the
cavitation region becomes almost zero and the pressure inside the
air increases as it is compressed. Velocity profiles indicate that
the velocity gradually increases towards smaller radii both in the
cavitation region and the $Pb$ shell, i.e. the inner edge of the
shell is moving faster than its outer edge. During early stages of
the collapse, the interface velocity (which is equal to the fluid
velocity at the position of the interface) roughly corresponds to
Eq.~\ref{eq:VelInt}, but later increases due to the converging
geometry.

If we look at the flow field structure inside the gas cavity one
can observe a shock wave propagating through it. This shock wave
is generated inside the air due to the sudden acceleration of the
interface. The interface is analogous to a piston at rest that
suddenly begins moving into a quiescent gas at constant velocity.
In this situation, a shock front immediately appears, moving away
from the piston with a constant supersonic speed. Ahead of the
shock front the gas is at rest, while behind the shock it moves at
the same velocity as the piston, i.e. the interface velocity in
our case (see ~\cite{CourantFri}  $\S 3$). Note that our numerical
method is not sufficient for a high-accuracy solution of shock
wave propagation inside the compressed gas. However, as mentioned
earlier, the gas dynamics has little effect on the collapse
therefore current numerical setup is sufficient for this study.

\begin{figure}[tbp]
\begin{center}
\includegraphics[width=1\linewidth]{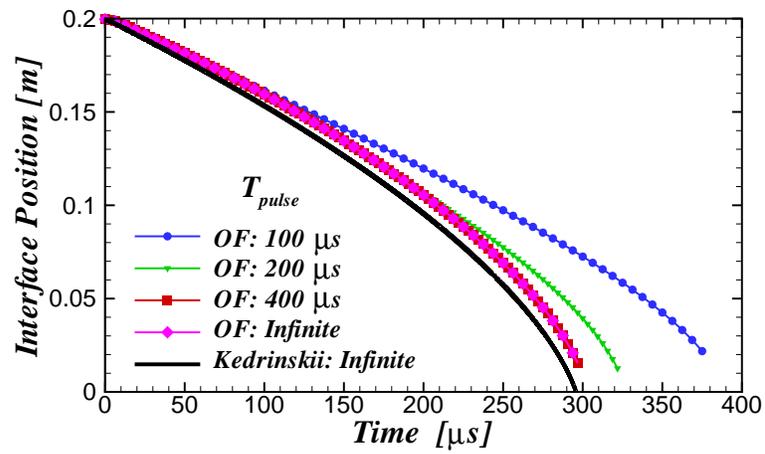}
\end{center}
\caption{Effect of pressure pulse duration $T_\mathrm{pulse}$ on motion
of liquid-gas interface during collapse of initially unperturbed cylindrical cavity.
Semi-analytical solution of Kedrinski~\cite{Kedrinskii} is also plotted as
black solid line. Maximum pulse pressure is $P_\mathrm{max}=1.5$ GPa.}
\label{LengthOfPulseRefined}
\end{figure}

It is necessary to accurately predict the trajectory of the
liquid-gas interface throughout the collapse so that the
compression efficiency of our system can be estimated. The motion
of an initially unperturbed interface in  cylindrical geometry is
shown in Fig.~\ref{LengthOfPulseRefined}. The four different lines
show our numerical results obtained for the pressure pulses of
various durations and with maximum pressure $P_\mathrm{max}=1.5$
GPa. The theoretical solution of Kedrinskii ($\S$ 1.4 in
~\cite{Kedrinskii}) is also shown by the black solid line for
comparison. One can see that the duration of the pulse influences
the collapse time; longer pulses compress the cavity faster. This
effect, however, diminishes as the pulse duration is increased,
such that no difference in collapse time is observed for pulses
with $T_\mathrm{pulse} \ge 400$ $\mu$s. Our results for the longer
pulses are also in a very good agreement with a theoretical
solution by Kedrinskii~\cite{Kedrinskii}) developed for studying
underwater explosions\footnote[1]{Detonating an explosive charge
underwater distributes energy between detonation products and
liquid. The gas in the explosive cavity is heated and acts as a
piston on the water, generating a shock wave. The Kirkwood-Bethe
approach~\cite{Cole} to the problems of underwater explosion can
be used to derive the pulsation equation, the equation of motion
for the edge of the cavity. Because it applies to states after the
detonation, it can also be applied to our problem of a shock
impinging on a pre-existing cavity.  The pulsation equation for a
one-dimensional isentropic compressible liquid flow is presented
by Kedrinskii~\cite{Kedrinskii} as
\begin{equation}
R\left(1-\frac{\dot{R}}{c}\right)\ddot{R}+\frac{3}{4}\nu\left(1-\frac{\dot{R}}{3c}\right)\dot{R}^{2}=\frac{\nu}{2}
\left(1+\frac{\dot{R}}{c}\right)H+\frac{R}{c}\left(1-\frac{\dot{R}}{c}\right)\frac{dH}{dt},
\end{equation}

where $R$ is the cavity radius, $c$ is the local speed of sound,
$H$ is the enthalpy on the cavity wall from the liquid side, and
$\nu$ depends on the symmetry, which can be planar ($\nu=0$),
cylindrical ($\nu=1$), or spherical ($\nu=2$). When the pressure
in the cavity is much less than the shock pressure, the enthalphy
at the interface is always zero ($H=0$), eliminating the RHS. Then
the liquid collapse is determined only by geometric convergence,
which can be solved numerically.}. Some additional results
concerning the effect of the pressure pulse amplitude as well as
collapse characteristics of the initially unperturbed spherical
cavity can be found in our earlier work~\cite{Suponitsky}.

 It is
worth noting that in the current numerical setup the gas never
becomes sufficiently pressurized to affect the trajectory of the
interface, which accelerates all the way to the axis due to
geometrical convergence. In reality, however, the interface
undergoes rapid deceleration during the very latest stages of
compression because the gas pressure becomes comparable to the
pressure in the $Pb$ shell. This deceleration is very important as
the interface becomes Rayleigh-Taylor unstable during this phase.

\subsection{The Richtmyer-Meshkov Instability} \label{sec:RM}

Now we turn our attention to the development of the RM instability
during the collapse due to imperfections that may be present on
the liquid-gas interface. In order to understand how various
perturbations are going to affect the compression efficiency of
our system, we study effects of the initial perturbation amplitude
and azimuthal mode number. The parameters for each simulation are
summarized in Table~\ref{tab:SimList}. In all cases, the pressure
pulse has an amplitude of $P_\mathrm{max}=1.5$ GPa and a duration
of $T_\mathrm{pulse}=100$ $\mu$s.

\begin{figure}[tbp]
\begin{center}
\includegraphics[width=1\linewidth]{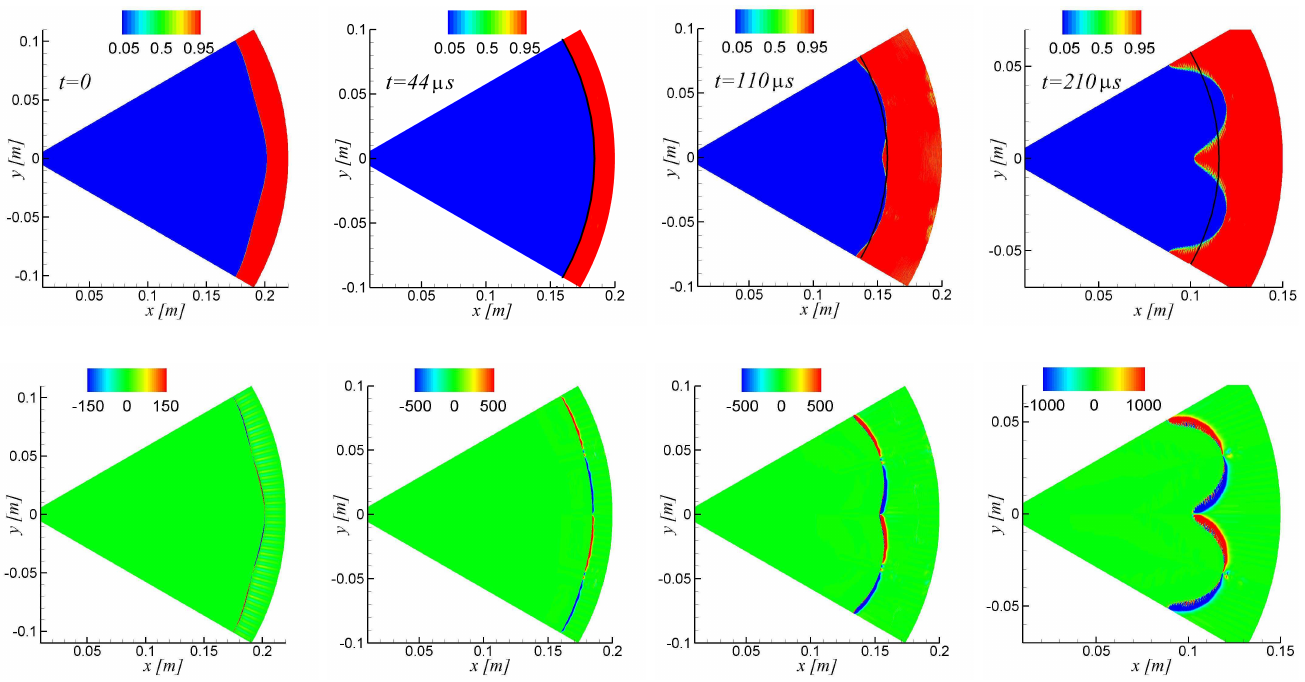}
\end{center}
\caption{\label{InitialEvolution} Early development of RM
instability shown by VoF contours (first row) and contours of the
vorticity component in $\hat{z}$ direction (units $[1/s]$)
multiplied by VoF (second row). Initial perturbation is at
azimuthal mode $n=12$ with amplitude of $h_{0}=2$ mm (Case N12A002
in Table~\ref{tab:SimList}). Black solid line shows interface
position for initially unperturbed case. Pressure pulse has
amplitude $P_\mathrm{max}=1.5$ GPa and duration
$P_\mathrm{pulse}=100$ $\mu$s.}
\end{figure}

\begin{figure}
\begin{center}
\includegraphics[width=1\linewidth]{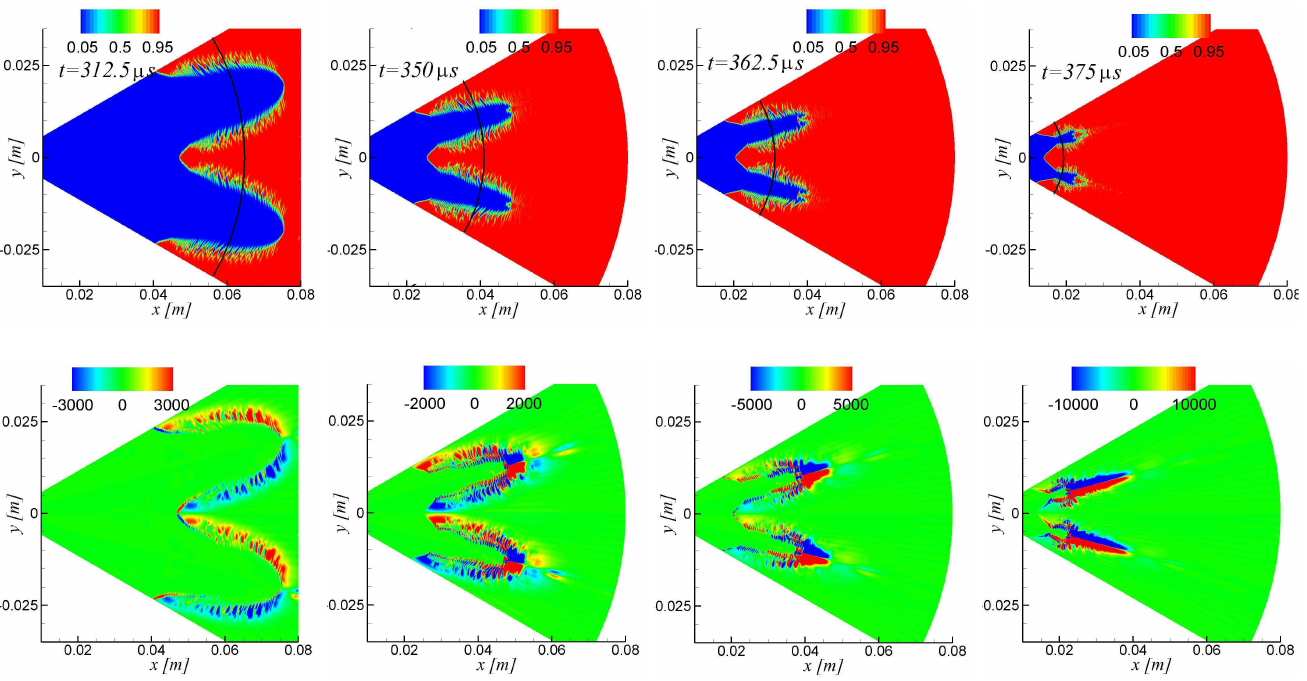}
\end{center}
\caption{\label{LateEvolution} Late development of RM instability
shown by VoF contours (first row) and contours of vorticity
component in $\hat{z}$ direction (units $[1/s]$) multiplied by VoF
(second row). Initial perturbation is at azimuthal mode $n=12$
with amplitude of $h_{0}=2$ mm (Case N12A002 in
Table~\ref{tab:SimList}). Black solid line shows interface
position for initially unperturbed case. Pressure pulse has
amplitude $P_\mathrm{max}=1.5$ GPa and duration
$P_\mathrm{pulse}=100$ $\mu$s.}
\end{figure}

A typical perturbation evolution during the early and late stages
of the collapse is shown in Figs.~\ref{InitialEvolution} and
~\ref{LateEvolution}, respectively. In both figures, the VoF
contours are plotted in the first row and the corresponding
contours of the $\hat{z}$ vorticity component multiplied by VoF
are plotted in the second row\footnote[2]{ Due to very high
velocities and gradients, vorticity attains very high values in
the gas, hiding what happens in the Pb. Multiplying vorticity by
VoF basically gives us vorticity contours only in the Pb, which is
of greatest interest.}.

The position of
the initially unperturbed interface is also shown by the black
solid line. Results are presented for case N12A002 listed in
Table~\ref{tab:SimList} with an
initial amplitude of $h_{0}=2$ mm and $n=12$.

One can see that once the pressure pulse interacts with the
perturbed liquid-gas interface (Fig.~\ref{InitialEvolution} at
$t=0$), vorticity is immediately generated in the vicinity of the
interface because of initial misalignment of density and pressure
gradients across the interface, i.e. the mechanism of baroclinic vorticity
generation. For a pulse passing from a heavy fluid
into a light one, the deposited vorticity initially acts in the
direction opposite to that of the perturbation,
smoothing the interface during the early evolution stages
(Fig.~\ref{InitialEvolution} at $t=44$ $\mu$s). Vorticity then
carries on to deflect the interface leading to the growth of the
perturbation in the opposite direction, i.e. phase inversion
(Fig.~\ref{InitialEvolution} at $t=110$ $\mu$s and
$t=210$ $\mu$s). The asymmetry between the spikes and
bubbles observed in Fig.~\ref{InitialEvolution} at
$t=210$ $\mu$s indicates that the perturbation is entering a
nonlinear stage of evolution.

There are two non-dimensional parameters that can be used to
characterize evolution of the perturbation amplitude. The first
one is the ratio of the perturbation amplitude and wavelength
$h(t)/\lambda (t)$. Similar to the planar case, perturbation
evolution is considered to be linear when $h(t)/\lambda (t) \ll
1$. However, in cylindrical geometry the wavelength of the
perturbation decreases as the cavity is compressed so that
nonlinear effects become prominent earlier. The second parameter
is the ratio between the disturbance amplitude and radius of the
cavity $h(t) / R(t)$. This parameter indicates how much the
perturbation evolution is influenced by the curvature of the
interface. For small amplitude initial disturbances parameter
$h(t)/R(t)$ is small.  At early stages of the collapse only low
azimuthal modes are expected to be influenced by the curvature of
the interface as they have significant ratios of $\lambda (t)
/R(t)$, whereas early evolution of the perturbations at higher
azimuthal modes is expected to be similar to that of the planar
case. As the cavity continues to be compressed, however, the
decrease in cavity radius increases the number of modes that are
affected by curvature. Therefore, while the initial motion may be
negligibly different from the planar case, we expect convergence
effects to manifest themselves at some point during the collapse.

Keeping the above in mind we follow the spike evolution in
Figs.~\ref{InitialEvolution} and \ref{LateEvolution}. One can see
that after phase inversion is complete ($t=110$, 210, and 312.5
$\mu$s), the spike amplitude grows, i.e. the distance increases
between the crest of the perturbation and the position of the
initially unperturbed interface. At later stages ($t=350$, 362.5,
and 375 $\mu$s), the spike amplitude starts to decrease. This
decrease in the perturbation amplitude correlates with the
increase of parameter $h(t)/R(t)$. By examining the vorticity
distribution, a reversal of the vorticity along the spike
interface near the crest can be observed. This change in direction
of rotation correlates with the direction of interface deflection,
best seen by comparing times $t=210$ and 312.5 $\mu$s. One can
also see the Kelvin-Helmholtz (K-H) instability that develops on
the sides of the bubbles and spikes at later times $t>300$ $\mu$s,
giving them a serrated appearance.

Another interesting phenomenon that can be observed in
Figs.~\ref{InitialEvolution} and \ref{LateEvolution} is the
formation of narrow molten lead jets (ribs) originating from the
back of the bubble during the latest stages of the collapse
($t=350$, 362.5, and 375 $\mu$s). Formation of such jets has been
observed in our simulations for perturbations with azimuthal mode
numbers higher than four $(n > 4)$. The prominence of the jets is
dependent on the amplitude and mode of the initial perturbation.
For this example case, the narrow jets only form but do not
overtake the original spikes during the collapse. Instead, the
spikes grew sufficiently to contact one another near the center,
despite the deceleration they experience late in the collapse.

\begin{figure}
\begin{center}
\includegraphics[width=1\linewidth]{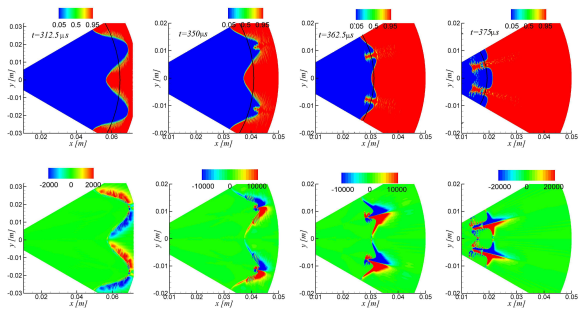}
\end{center}
\caption{\label{RibJet} Formation of rib-like jets during late
stages of collapse. VoF contours are shown in first row. Contours
of vorticity component in $\hat{z}$ direction (units $[1/s]$)
multiplied by VoF are shown in second row. Initial perturbation is
mode $n=12$ with amplitude $h_{0}=1$ mm (Case N12A001 in
Table~\ref{tab:SimList}). Black solid line shows interface
position for initially unperturbed case. Pressure pulse has
amplitude $P_\mathrm{max}=1.5$ GPa and duration
$T_\mathrm{pulse}=100$ $\mu$s.}
\end{figure}

For the case of a perturbation with the same mode, but a lower
initial amplitude, the situation is different, as shown in
Fig.~\ref{RibJet}. This figure is in the same format as
Fig.~\ref{LateEvolution}, but for case N12A001 in
Table~\ref{tab:SimList}, which has a lower initial amplitude of
$h_{0}=1$ mm. In this case these narrow molten lead jets move fast
enough to overtake the original spikes and reach the center first.
The results indicate that these narrow jets are also formed by
redistribution of vorticity. We are not aware of such jets being
observed in other works that use two gases with a moderate Atwood
number as the working fluids. A very similar phenomenon has been
observed, however, in the recent work of Enriquez \emph{et
al.}~\cite{Gekle}, in which an air cavity formed by collision of a
solid body with a liquid reservoir collapses due to hydrostatic
pressure (see Fig. 1 in~\cite{Gekle}). The overall collapse
process described in~\cite{Gekle} is remarkably similar to the one
obtained by our simulations.

\subsubsection{Effect of Initial Perturbation Amplitude}
\label{sec:InitAmp} 
\begin{figure}
\begin{center}
\includegraphics[width=1\linewidth]{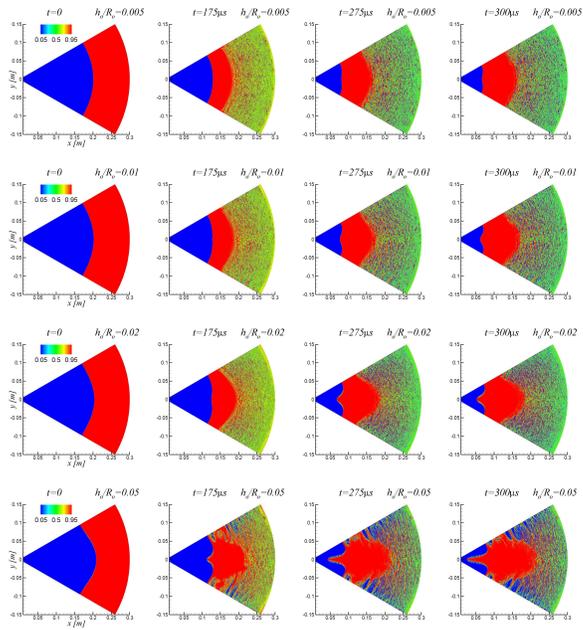}
\end{center}
\caption{\label{AmpN6vof} Effect of initial perturbation amplitude
shown by VoF contours. Perturbation is mode $n=6$ and pressure
pulse is $P_\mathrm{max}=1.5$ GPa. Each row of figure corresponds
to a unique initial amplitude: $h_{0}=1$, 2, 4, and 10 mm; Cases
N6A001, N6A002, N6A004, and N6A010 in Table~\ref{tab:SimList}.}
\end{figure}

The evolution of perturbations with various
initial amplitudes is shown in Fig.~\ref{AmpN6vof} by VoF
contours. Rows in the figure correspond to the evolution of $n=6$
perturbations with different initial amplitudes: cases N6A001,
N6A002, N6A004 and N6A010 in Table~\ref{tab:SimList}. One can see
that for the small amplitude initial perturbation (first row) no
significant nonlinear effects are observed and the spikes and
bubbles remain nearly symmetric throughout the time period shown.
As the amplitude increases, nonlinear effects begin to manifest
themselves in the growing asymmetry between the spikes and
bubbles. The spike appears to accelerate and becomes sharper,
whereas the bubble appears to stagnate. For the current Atwood
number of $A \approx -1$, the spikes are significantly sharper
than those simulated for lower Atwood numbers. This has also been
observed by Tian \emph{et al.}~\cite{tian}. In addition,
Fig.~\ref{AmpN6vof} illustrates that the shape of the molten lead
shell surrounding the gas cavity is affected by the initial
imperfections of the interface. The distortion of the shell
increases as the initial perturbation amplitude is increased.  For
the largest tested amplitude (row four), the thickness of the
shell behind the bubble almost goes to zero.

Before proceeding to the plots of the evolution of spikes and
bubbles, we would like to once more clarify the notation being
used in all our plots. We follow extrema of the perturbation
throughout the entire simulation, therefore our notation of
`spike' and `bubble' corresponds to that usually used in the
literature from the moment the phase inversion has occurred, as
explained earlier in the validation section.

\begin{figure}
\begin{center}
\includegraphics[width=1\linewidth]{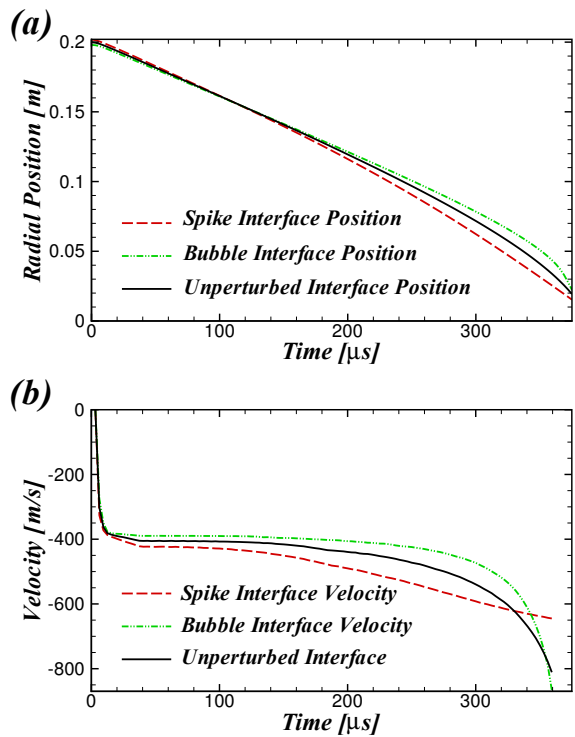}
\end{center}
\caption{\label{TypicalIntEvol}(a) Typical position of spike
interface (red broken line), bubble interface (green dashed-dotted
line), and initially unperturbed interface (black solid line); (b)
Typical velocity of spike interface (red broken line), bubble
interface (green dashed-dotted line), and initially unperturbed
interface (black solid line). Case N6A002 in
Table~\ref{tab:SimList}, pulse pressure $P_\mathrm{max}=1.5$ GPa,
pulse length $T_\mathrm{pulse}=100$ $\mu$s.}
\end{figure}

This is illustrated in Fig.~\ref{TypicalIntEvol} for the case
N6A002 in Table~\ref{tab:SimList}. Part (a) shows a typical
evolution of the spike (red broken line)  and bubble (green
dash-dot line) interface position along with the position of the
initially unperturbed interface (black solid line). Part (b) of
the figure shows the corresponding interface velocities. Time
$t=0$ corresponds to the moment when the pressure pulse hits the
interface and the collapse begins. One can see that at $t=0$, the
red and green lines corresponding to the maximum and minimum of
the initial perturbation are above and below the radial position
of the initially unperturbed interface (black line), respectively,
and the difference between those lines defines the amplitude of
the initial perturbation $h_{0}$. The perturbation decreases in
amplitude until around $t \approx 100$ $\mu$s  (when phase
inversion occurs) and then starts to grow in the opposite
direction. From that moment our notation of 'spike' and 'bubble'
matches that commonly used in literature, i.e. a finger of light
fluid poking into heavy fluid for a `bubble', and that of heavy
into light for a `spike'.

At late stages of the collapse the difference between red and
black lines as well as between green and  black lines starts to
decrease again eventually accompanied by another reversal, which
indicates formation of the narrow molten lead jets.
Fig.~\ref{TypicalIntEvol} (b) shows the rapid acceleration of the
interface resulting from its interaction with the pressure pulse.
During early stages of the collapse, the velocities of the spikes,
bubbles, and that of the unperturbed interface are nearly
constant. Later the velocity of the unperturbed interface
increases considerably due to the geometric convergence.

\begin{figure}
\begin{center}
\includegraphics[width=1\linewidth]{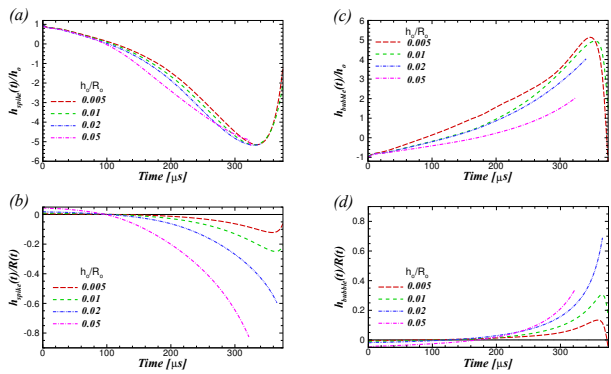}
\end{center}
\caption{\label{AmpN6} Effect of initial amplitude on
perturbation growth. Left and right columns correspond to
spike and bubble evolution. (a,c) Evolution of normalized
perturbation amplitude; (b,d) Evolution of ratio between
perturbation amplitude and radius of unperturbed gas cavity;
 Azimuthal mode number $n=6$, pressure pulse with $P_{max}=1.5$ GPa and
$T_{pulse}=100$ $\mu$s.}
\end{figure}

The effect of initial amplitude on the perturbation growth is
shown in Fig.~\ref{AmpN6} for cases N6A001, N6A002, N6A004 and
N6A010 in Table~\ref{tab:SimList}. The left column (parts a and b)
and the right column (parts c and d) in Fig.~\ref{AmpN6} show the
amplitude evolution of the spikes and bubbles, respectively. The
top row shows the amplitude normalized by its initial value
$h_{0}$, while the bottom row shows the amplitude normalized by
the radius of the unperturbed gas cavity $R(t)$.

Examining the growth characteristics of the spike we can observe
the following: (i) at early times growth of the spikes scales well
with the initial perturbation amplitude for all amplitudes under
consideration, (ii) after the phase inversion when the curves
cross zero for the first time, spike amplitude growth is faster
for higher initial amplitudes, (iii) at large initial amplitudes
the spike arrives at the center while it is still growing, so that
no decrease of the spike amplitude is observed during the latest
stages. For small initial amplitudes the spikes experience
deceleration during late times, leading to a rapid decrease in
spike amplitude. By comparing the growth characteristics of spikes
and bubbles, it is apparent that the bubble amplitude does not
scale as well with the initial perturbation amplitude, even early
in the collapse. The bubble amplitude growth is significantly
reduced for larger initial perturbations when compared to smaller
ones. For the small amplitude perturbations, a decrease in bubble
amplitude can be seen at the latest stages. This decrease is
related to formation of the rib-like jets and their rapid
propagation toward the center of the cavity, as discussed earlier.

\begin{figure}
\begin{center}
\includegraphics[width=1\linewidth]{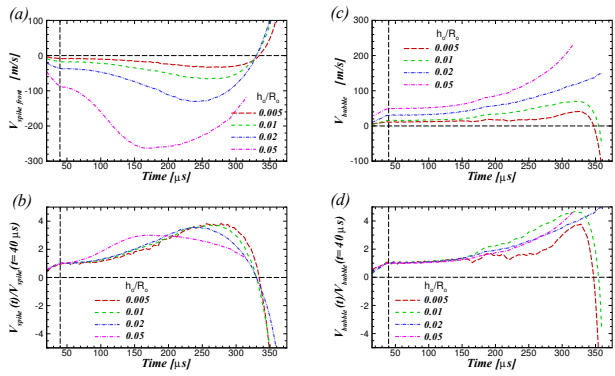}
\end{center}
\caption{\label{VelN6} Effect of initial amplitude on
perturbation growth rate. Left and right columns correspond to
spike and bubble evolution, respectively. (a,c) Perturbation growth rate;
(b,d) Perturbation growth rate normalized by its initial
growth rate at $t=40$ $\mu$s.
 Azimuthal mode number $n=6$, pulse pressure $P_\mathrm{max}=1.5$ GPa, pulse length
$T_\mathrm{pulse}=100$ $\mu$s.}
\end{figure}

The effect of initial amplitude on the growth rates of spikes and
bubbles is shown in Fig.~\ref{VelN6}. The left and right columns
of the figure show the velocities of the spikes and bubbles
(relative to the velocity of the unperturbed interface)
corresponding to the data in Fig.~\ref{AmpN6}. Dimensional
velocities are plotted in the first row of the figure and the same
velocities normalized by the corresponding velocity at $t=40$
$\mu$s (immediately after the initial acceleration of the
interface has been completed) are plotted in the second row. In
Figs.~\ref{AmpN6}(a) and~\ref{AmpN6}(c) a negative velocity
corresponds to the situation in which the perturbed interface
(either spike or bubble) moves inwards faster than the initially
unperturbed interface, whereas a positive velocity indicates that
the perturbed interface moves inwards slower than the unperturbed
interface (although it still moves inwards).

From the velocity plots one can see that after some finite initial
time required to accelerate the interface from rest ($t \approx
40$ $\mu$s), the velocities of both spikes and bubbles approach a
nearly constant value for a little while. This value is taken as
the initial velocity that is used for scaling. Both spikes and
bubbles undergo gradual acceleration until late times, when there
is rapid deceleration, except for cases with large initial
amplitudes in which the spikes reach the axis before this
decelerate phase can occur. For all amplitudes under
consideration, the bubble growth rate scales well with initial
bubble velocity until the collapse is well underway. The spike
growth rate also scales well, except for large amplitude
perturbations, in which the growth rate saturates more quickly
than the other cases.

\subsubsection{Effect of Azimuthal Mode Number}
\label{sec:AziModNum}

\begin{figure}
\begin{center}
\includegraphics[width=1\linewidth]{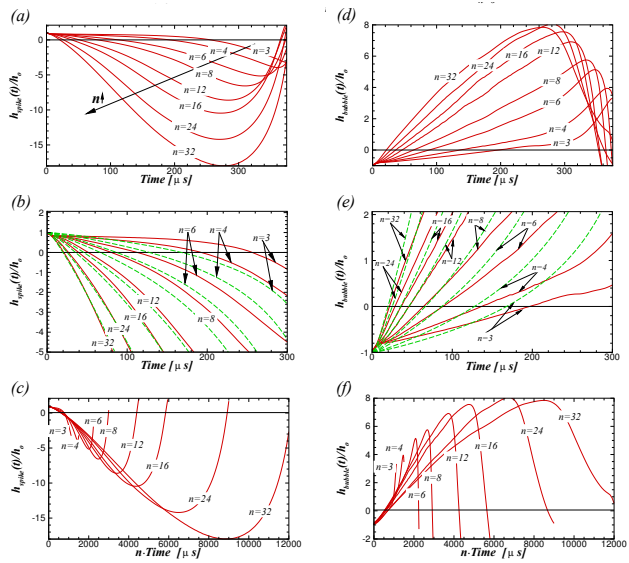}
\end{center}
\caption{\label{Aziho} Effect of the azimuthal mode number $n$ on
perturbation growth. Left and right columns correspond to
spike and bubble evolution, respectively. (a,d) Normalized perturbation
amplitude; (b,e) zoom of (a,d) (red solid lines)
along with small amplitude theoretical model of
Mikaelian~\cite{Mikaelian} (green broken lines); (c,f) time
scaled by mode $n$. Pressure pulse with
$P_\mathrm{max}=1.5$ GPa and $T_\mathrm{pulse}=100$ $\mu$s. }
\end{figure}

The behaviour of spikes and bubbles is tested for various
representative azimuthal mode numbers in the range $3 \le n \le
32$, with the results displayed in Fig.~\ref{Aziho}. The left
columns (parts a, b, and c) and the right columns (parts d, e, and
f) correspond to the evolution of spikes and bubbles,
respectively. The first row of the figure (parts a and d) shows
amplitude evolution of spikes and bubbles normalized by the
initial perturbation amplitude $h_{0}$. The second row (parts b
and e) shows a zoomed-in section of the first row plot together
with the theoretical model for the small amplitude perturbations
of Mikaelian~\cite{Mikaelian} (green dashed lines). Finally, the
third row (parts c and f) show the same data when time is scaled
by the mode $n$. All simulations have been carried out for an
initial perturbation amplitude of $h_{0}/R_{0}=0.005$.

\begin{figure}
\begin{center}
\includegraphics[width=1\linewidth]{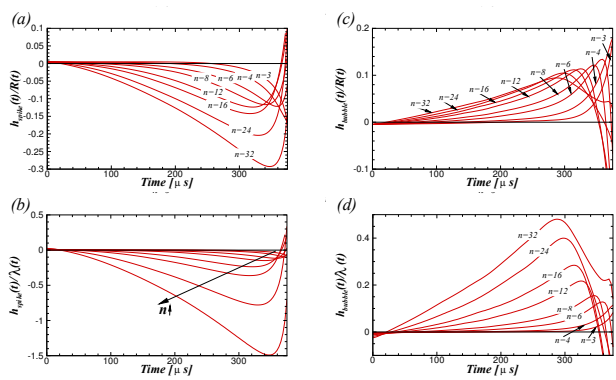}
\end{center}
\caption{\label{AziRLambda} Effect of azimuthal mode number $n$ on
perturbation evolution. Left and right columns correspond to the
spike and bubble evolution, respectively. (a,c) Evolution of ratio
between perturbation amplitude and radius of unperturbed gas
cavity; (b,d) Evolution of the ratio between perturbation
amplitude and wavelength. Pressure pulse with $P_\mathrm{max}=1.5$
GPa and $T_\mathrm{pulse}=100$ $\mu$s. }
\end{figure}

From the plots shown in Fig.~\ref{Aziho} the following can be
observed: (i) for the setup and parameters under consideration the
perturbation evolution at low azimuthal numbers ($n=3,4$) differs
from that at higher azimuthal mode numbers. In particular, the
decrease in the perturbation amplitude (and in most cases a second
phase inversion) observed at higher azimuthal modes at the latest
stages of the collapse does not occur at low azimuthal mode
numbers; (ii) at higher azimuthal modes ($n=24,32$) the maximum
amplitude attained by spikes is significantly higher than that of
bubbles; (iii) the initial evolution of spikes agrees well with
Mikaelian's theoretical model~\cite{Mikaelian} for perturbations
with large $n$. The predictions of the theoretical model are less
favorable for bubble evolution; (iv) the phase inversion time
roughly scales with the azimuthal mode number; (v) the second
phase inversion, due to formation of the narrow (rib-like) jets at
the head of each bubble, is clearly seen in parts (c) and (f) of
the figure. However, in the case of large azimuthal mode, the
cavity collapses before the second phase inversion is completed,
as can be seen from the $n=32$ curves. Fig.~\ref{AziRLambda} shows
the same results as in Fig.~\ref{Aziho}, but with the perturbation
amplitude normalized by the cavity radius $R(t)$ (first row) and
perturbation wave length $\lambda(t)$ (second row). The left and
right columns of the figure correspond to the spike and bubble
evolution, respectively. One can see that at later stages of the
collapse, both ratios, $h(t)/R(t)$ and $h(t)/\lambda(t)$ attain
significant values for all perturbations being considered. This
implies that both the nonlinear effects and the effect of the
interface curvature become important for all perturbations at some
point during the collapse.

\begin{figure}
\begin{center}
\includegraphics[width=1\linewidth]{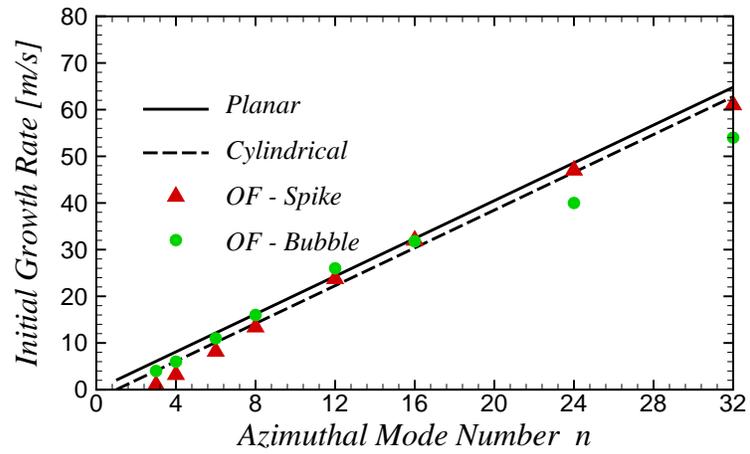}
\caption{\label{LinVelDataComp} Effect of azimuthal mode number on
perturbation early growth rate. Red triangles and green circles
correspond to initial growth rates of spikes and bubbles,
respectively. Solid line is planar model of
Richtmyer~\cite{richtmyer} (Eq.~\ref{eq:planar}). Dashed line is
cylindrical model of Mikaelian~\cite{Mikaelian}
(Eq.~\ref{eq:cyl}). Pressure pulse with $P_\mathrm{max}=1.5$ GPa
and $T_\mathrm{pulse}=100$ $\mu$s.}
\end{center}
\end{figure}

Finally, Fig.~\ref{LinVelDataComp} shows the effect of the
azimuthal mode number on the initial perturbation growth rate,
taken at $t \approx 40$ $\mu$s, after the initial
acceleration of the interface has been completed.
The growth rates of spikes and bubbles
are shown by red triangles and green circles,
respectively. Here we reiterate that these initial growth
rates are calculated before any of the azimuthal modes have
completed phase inversion. Therefore, the `spikes' and `bubbles'
referred to in the legend of Fig.~\ref{LinVelDataComp} are outward
and inward bulges, respectively.

The data is fitted to two linear models.  The first is that
of Richtmyer~\cite{richtmyer} for planar geometry,
\begin{equation}
\dot{h}_\mathrm{planar}=h_{0} k  A V_\mathrm{interface},
\label{eq:planar}
\end{equation}
where as usual, $h_{0}$ denotes the initial perturbation
amplitude, $A$ is an Atwood number, $k=2\pi/\lambda=n/R_{0}$ is
the wave number, $n$ is the azimuthal mode, and
$V_\mathrm{interface}$ is the initial velocity of the undisturbed
interface (see Eq.~\ref{eq:VelInt}). The second model is that of
Mikaelian~\cite{Mikaelian} for cylindrical geometry,
\begin{equation}
\dot{h}_\mathrm{cylindrical}=(n A-1)\frac{h_{0}}{R_{0}} V_\mathrm{interface},
\label{eq:cyl}
\end{equation}
where $R_{o}$ is the initial radial position of the interface.
The growth rate given by the planar model is
shown by the solid line, while that of the cylindrical model is
shown by the dashed line.

The growth rates presented in Fig.~\ref{LinVelDataComp} show
encouraging agreement with linear models for the range of
perturbations being considered. It can be also seen that there is
a change in the growth rates pattern for the higher azimuthal
modes ($n > 16$ based of the our data set). This is probably
because of more pronounced nonlinear effects at those modes: in
all our simulations the parameter $h_{o}/R_{o}$ has been kept
constant and therefore, the parameter $h_{o}/\lambda_{o}$
increases with the increase in the azimuthal mode number of the
perturbation. One can also see some mismatch in the initial growth
rates of spikes and bubbles at lower azimuthal modes. This
disparity becomes less pronounced as azimuthal mode of the
perturbation increases (up to $n=16$).

\section{Summary}
\label{sec:Sum} In this work, the behaviour of the
Richtmyer-Meshkov instability was studied for the case of a
cylindrical gas (air) bubble compressed by an imploding molten
lead shell. The main contribution of this work is to explore the
RM instability in the extreme regime of Atwood number $A=-1$ with
a liquid (molten lead) as one of a working fluids. Our motivation
is to estimate the minimum smoothness required to achieve
efficient compression of the gas cavity. Simulations have been
performed using OpenFOAM software for a set of parameters relevant
to the prototype compression system under development at General
Fusion Inc. as a driver for magnetized target fusion. The main
results and conclusions are summarized below:

\begin{itemize}
    \item  In the regime of Atwood number $A=-1$, there is a disparity between the growth rates of spikes and bubbles; spikes undergo acceleration
    while bubbles move at nearly constant velocity. This disparity in growth rates becomes more prominent as the amplitude of the initial
    perturbation is increased.

    \item  The shape of the spikes obtained for the current set of parameters is different
    from that usually observed in the regime of moderate Atwood numbers.
    The spikes retain a sharp point with the Kelvin-Helmholtz instability
    producing serrated sides; they do not develop into the typical mushroom shape.

    \item During the latest stages of the collapse, when the non-dimensional
    parameters $\lambda (t) /R(t)$ and $h(t)/R(t)$
    are no longer small, the spike amplitude starts
    to decrease. For some range of perturbation azimuthal modes and amplitudes, this is the onset of a second phase inversion.

    \item The formation of narrow molten lead jets propagating inwards and originating from the top of the bubbles has been observed during
    the latest stages of the collapse for modes $n>4$. To the best of our knowledge these jets have not been observed at a gas-gas interface
    with moderate Atwood number.

    \item To maintain sufficient compression efficiency, low-mode interface perturbations are not likely to  be detrimental.  However,
    high-mode perturbations are problematic and must be kept to a minimum.

\end{itemize}

This numerical setup seems to produce valuable results despite its
lack of sophisticated modeling for all physical phenomena
involved. It will be interesting to explore the effect of
incorporating more physics into the simulation. There is also
opportunity to further examine the effects of cavitation model,
shock wave capturing scheme, rotation of the fluid and magnetic
field on the dynamics of the gas cavity collapse.


\begin{table}
\caption{Planar Case: Initial Growth Rate } \label{tab:RecRes}
\centering
\begin{tabular}{|c|c|c|c|}
  \hline
  Case & $h_{0} [m]$ & $T_\mathrm{pulse}\, [\mu s]$ & $\dot{h}\, [m/s] $  \\
  \hline
  Richtmyer model~\cite{richtmyer} (Eq.~\ref{eq:Richt}) & $0.002$ & infinite & $20.7 $ \\
  OpenFOAM simulation     & $0.002$ & infinite & $23.0$    \\
  OpenFOAM simulation     & $0.002$ & $100$    & $9.8$   \\
  \hline
\end{tabular}
\vspace{10pt}
\caption{List of parameters for the simulations
performed}\label{tab:SimList} \centering
\begin{tabular}{|c|c|c|c|c|c|}
\hline
  No. & Name & n & $h_{0}$ [m] & $h_{0}/R_{0}$ & $h_{0}/\lambda_{0}$ \\
  \hline
  1 & N3A001 & 3 & 0.001 & 0.005 & 0.002387 \\
  \hline
  2 & N4A001 & 4 & 0.001 & 0.005 & 0.003183 \\
  \hline
  3 & N6A001 & 6 & 0.001 & 0.005 & 0.004775 \\
  \hline
  4 & N8A001 & 8 & 0.001 & 0.005 & 0.006366 \\
  \hline
  5 & N12A001 & 12 & 0.001 & 0.005 & 0.009549 \\
  \hline
  6 & N12A002 & 12 & 0.002 & 0.010 & 0.019098 \\
  \hline
  7 & N16A001 & 16 & 0.001 & 0.005 & 0.01273 \\
  \hline
  8 & N24A001 & 24 & 0.001 & 0.005 & 0.01909 \\
  \hline
  9 & N32A001 & 32 & 0.001 & 0.005 & 0.02546 \\
  \hline
  \hline
  10 & N6A002  & 6 & 0.002 & 0.010 & 0.009549 \\
  \hline
  11 & N6A004 & 6 & 0.004 & 0.020 & 0.01909 \\
  \hline
  12 & N6A010 & 6 & 0.010 & 0.050 & 0.04775 \\
  \hline
  \hline
  13 & N3A004 & 3 & 0.004 & 0.020 & 0.009549 \\
  \hline
  14 & N3A008 & 3 & 0.008 & 0.040 & 0.01909 \\
  \hline
  \hline
\end{tabular}
\end{table}

\newpage
\bibliographystyle{elsarticle-num}

\newpage







\vspace{-10pt}


%
%
%
%
%
%

\end{document}